\documentclass[twocolumn]{aastex701}
\usepackage{amsmath}
\usepackage{subfigure}
\usepackage{xcolor}
\usepackage{float,color}
\usepackage{gensymb}

\usepackage{graphicx}
\usepackage{txfonts}
\usepackage{caption}
\usepackage{enumitem}
\usepackage{hyperref} % for clickable links
\hypersetup{
    colorlinks=True,
    linkcolor=blue,
    filecolor=magenta,
    urlcolor=cyan,
    citecolor=blue,
    pdfpagemode=FullScreen}

\usepackage{mdframed}

\begin{document}

\title{What do gravitational-wave observations tell us about Luminous Red Novae?}

\author[orcid=0009-0004-7242-4301]{Dhruv Jain}
\affiliation{Aryabhatta Research Institute of Observational Sciences (ARIES), Manora Peak, Nainital-263002, India}
\affiliation{Department of Applied Physics, Mahatma Jyotibha Phule Rohilkhand University, Bareilly-243006, India}
\email[show]{dhruvjain@aries.res.in}

\author[orcid=0000-0001-5318-1253]{Shasvath J. Kapadia} 
\affiliation{Inter-University Centre for Astronomy and Astrophysics, Post Bag 4, Ganeshkhind, Pune - 411007, India}
\email{shasvath.j.kapadia@gmail.com}

\correspondingauthor{Kuntal Misra}
\email{kuntal@aries.res.in}

\author[orcid=0000-0003-1637-267X]{Kuntal Misra} 
\affiliation{Aryabhatta Research Institute of Observational Sciences (ARIES), Manora Peak, Nainital-263002, India}
\email{kuntal@aries.res.in}

\author[orcid=0000-0001-9868-9042]{Dimple} 
\affiliation{Institute for Gravitational Wave Astronomy and School of Physics and Astronomy, University of Birmingham, Birmingham, B15 2TT}
\email{d.dimple@bham.ac.uk}

\author[orcid=0000-0001-9407-9845]{L. Resmi} 
\affiliation{Indian Institute of Space Science $\&$ Technology, Trivandrum 695547, India}
\email{l.resmi@gmail.com}

\author{Ajay Kumar Singh} 
\affiliation{Department of Applied Physics/ Physics, Bareilly College, Mahatma Jyotibha Phule Rohilkhand University, Bareilly-243001, India}
\email{}

\author[orcid=0000-0002-6960-8538]{K. G. Arun} 
\affiliation{Chennai Mathematical Institute, Siruseri, 603103 Tamil Nadu, India}
\email{kgarun@cmi.ac.in}


\begin{abstract}

Luminous Red Novae (LRNe) have been argued to be related to the ejection of common envelopes (CEs) in binary star systems. Ejection of CEs leads to tightened stellar orbits capable of forming compact binaries that merge in Hubble time. As these mergers are seen by gravitational-wave (GW) detectors such as LIGO, Virgo and KAGRA (LVK), we ask what the merger rates of compact binaries in LVK tell us about the fraction of LRNe that lead to the formation of compact binaries that merge in Hubble time. Using the observed volumetric rates of LRNe from the Zwicky Transient Facility (ZTF) and of compact binary mergers from LVK observations, we derive limits on the fraction of LRNe that produce compact binaries that merge in Hubble time. Assuming the LRNe rate closely follows the star formation rate at any redshift, we use the delay time distribution models for compact binaries to compute the compact binary merger rate. A comparison of this merger rate with the latest volumetric rates of compact binary mergers from the fourth GW transient catalog (GWTC-4) at the present epoch of LVK allows us to constrain the above fraction. We find that only a fraction as small as $\sim 10^{-3}$ (median) of the LRNe correspond to the GW-observed binary neutron star (BNS) and neutron star-black hole (NSBH) mergers. This potentially implies that the majority of the LRNe population will not lead to mergers of compact objects, but other end products, such as stellar mergers. 

\end{abstract}



\section{Introduction} \label{sec:intro}

The common-envelope evolution (CEE) phase of a binary star system is the phase in which two stars orbit each other within a common gaseous envelope produced by unstable mass transfer \citep{Paczynski1976}. The drag exerted by the gas on the binary leads to rapid orbital decay, producing tight binaries as end products of this evolution. Such tight binaries can potentially produce a compact binary system, such as a binary neutron star (BNS) or a neutron star-black hole (NSBH) \citep[see, e.g., ][for reviews]{Ivanova2013, Postnov2014}. However, due to the complexity of the underlying physics and limited observations, the theoretical understanding of CEE remains an open problem in modern astronomy. It is, therefore, essential that we utilise all available observational data to comprehend this intriguing phase of binary evolution. While there are observations of close binary systems that can be explained only if one invokes a CEE, the most direct signature of this phase is perhaps the Luminous Red Novae \citep[LRNe;][]{kasliwal2011,pastorello2019}. These are a class of transients whose absolute magnitudes lie between novae and supernovae ($-16\leq M_V\leq -10$), which are widely believed to be associated with the CEE of a binary star system. 

The most likely outcomes of a CEE are the merger of the two stars, or a tighter stellar binary via ejection of the common envelope \citep[CE;][]{Politano2010, tylenda2005, pastorello2019,Howitt2019, karambelkar2023}. Observational evidence for merger-induced LRNe is most compellingly provided by V1309 Scorpii, whose pre-outburst photometric evolution directly revealed the inspiral and subsequent merger of a contact binary system \citep{tylenda2011}. If the CEE leads to the formation of a tighter stellar binary, it can further evolve to form a BNS or NSBH binary. Mergers of such compact binaries in the nearby universe are detectable by the current-generation gravitational-wave (GW) interferometers LIGO, Virgo and KAGRA \citep[LVK;][]{aligo,avirgo,KAGRA}. In this Letter, we examine how GW observations, in conjunction with LRNe,  can provide a crucial handle on this phenomenon.

The advent of GW astronomy, marked by the landmark detection of GW150914 by the LVK Collaboration \citep{GW150914} and followed by the discoveries of BNSs~\citep{GW170817, GW190425} and NSBHs~\citep{LVK-NSBH}, have revealed a population of compact binary coalescences (CBCs) previously inaccessible to electromagnetic (EM) observations. Three observing runs of the LVK detector network have been completed, and the fourth one is currently ongoing. The most recent GW Transient Catalog~\citep[GWTC-4;][]{GWTC-4}, reports 218 CBCs with probability of astrophysical origin \citep{FGMC, Kapadia2019} $p_{\mathrm{astro}}\geq 0.5$. Though most of them are binary black holes (BBHs), the duration of science runs and the detections of a few BNS and NSBH events allow one to compute the volumetric rates of their mergers~\citep{GWTC-4Rates}. The corresponding local merger rates densities were estimated to be %\rho_{\rm BBH} \sim 14-26 ~\mathrm{Gpc^3} ~\mathrm{yr^{-1}}$,
 $\rho_{\rm NSBH} \sim 10-43 ~\mathrm{Gpc^{-3}} ~\mathrm{yr^{-1}}$, and $\rho_{\rm BNS} \sim 22-248 ~\mathrm{Gpc^{-3}} ~\mathrm{yr^{-1}}$ for NSBHs and BNSs \citep{GWTC-4Rates}, respectively. 

Despite reasonably precise rate estimates, due to their smaller number, population inferences are not yet efficient to unravel their formation channels. Theoretical studies show that BNS and NSBH mergers are more likely to be associated with galactic fields compared to environments such as dense star clusters~\citep{Belczynski2017, Ye2019}. CE evolution is a crucial step in the binary evolution that hardens the binary, enabling the subsequently formed compact binary to merge within Hubble time~\citep{Dominik2012, Ivanova2013,Olejak2021,Postnov2014}. Therefore, if these CBCs observed in GWs are formed via isolated binary evolution, they are likely to have a phase of CE ejection marking the end of CEE and would have produced an associated LRNe. Hence, the rate of LRNe and the rate of BNS and NSBH mergers should be closely related.

In this Letter, we explore this relation and put forward an astrophysically motivated model that relates the rates of LRNe and mergers of BNS and NSBH systems. We then calculate the fraction of LRNe associated with CBCs and discuss the potential implications of this fraction. Given the possibility of multiple formation channels at play for BBHs, making a one-to-one correspondence between their rate and that of the LRNe is difficult (see \cite{Howitt2019} for a discussion of binary black holes associated with LRNe). Therefore, we restrict our attention only to BNS and NSBH mergers in this work, following the widely held assumption that these are formed via isolated evolution.

\section{Relating luminous red novae to compact binary coalescences}

\begin{figure*}[htb]
    \centering
    \includegraphics[width=0.9\linewidth]{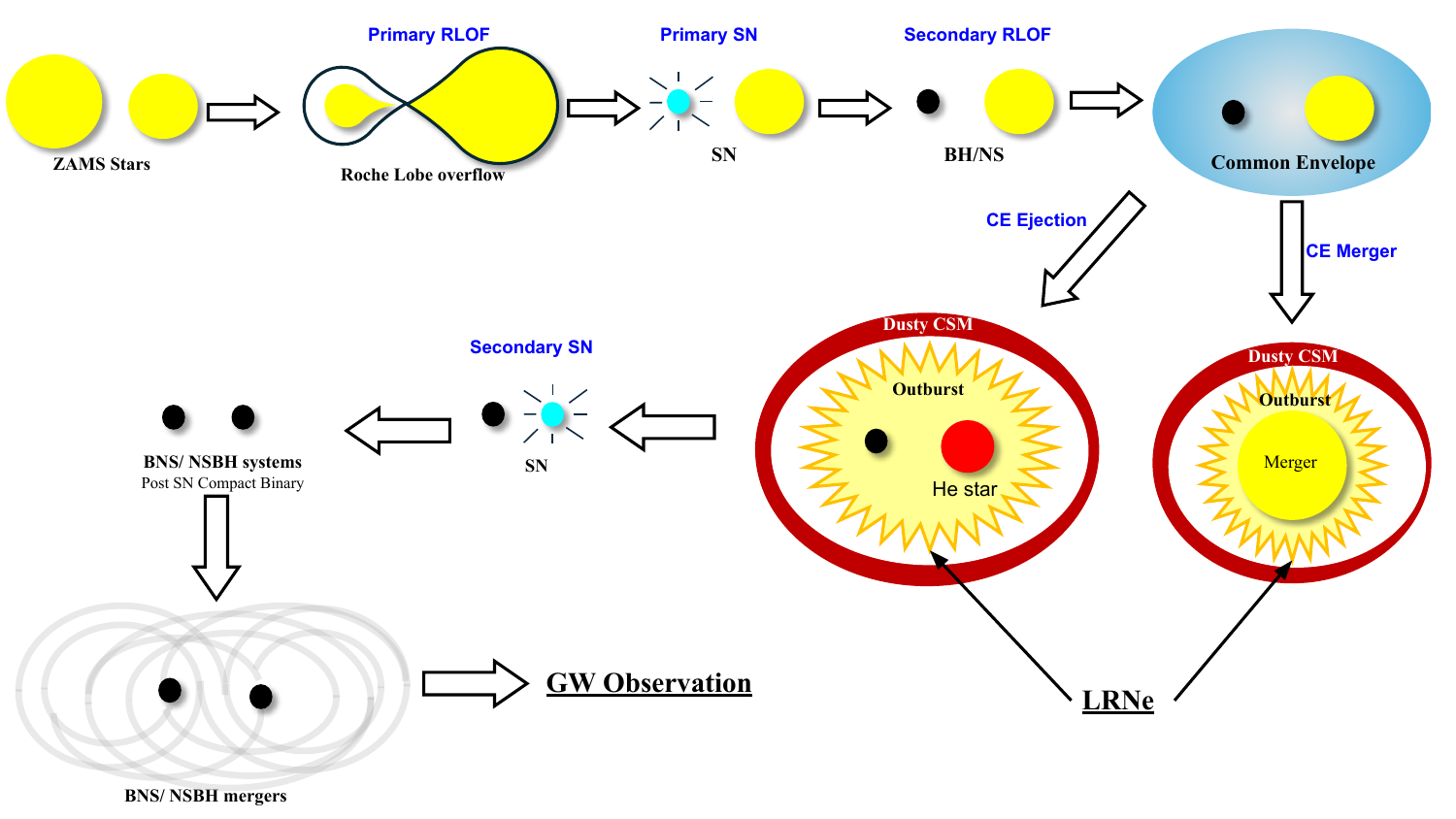}                   
    \caption{Schematic illustration of close-binary evolutionary pathways leading to LRNe and gravitational-wave (GW) progenitors. The common-envelope (CE) interaction produces an LRN and a dusty circumstellar medium. Successful CE ejection leaves a surviving binary, whereas complete merger yields a single remnant. Systems that survive the CE evolve into tighter orbits, may undergo stripped-envelope supernovae, and can form compact-object binaries (BNSs or NSBHs) that later coalesce as GW sources.}
    \label{fig:binary evo}
\end{figure*}

One of the primary pathways for forming BNS or NSBH systems is through the isolated binary evolution (IBE) channel. A standard evolutionary scenario, based on previous theoretical studies, is summarized in Figure~\ref{fig:binary evo}. A binary consisting of massive stars with masses of $\sim8$--$25,\mathrm{M_{\odot}}$ evolves such that the donor (primary) star loses its hydrogen-rich envelope via Roche-lobe overflow (RLO) to the secondary star, becoming a helium star. Systems that remain bound after the first supernova (SN) explosion \citep{yoon2010} are observed as high-mass X-ray binaries (HMXBs) or radio pulsars with OB-star companions \citep{johnston1992, kaspi1994}.

As the secondary star evolves and begins Roche-lobe overflow (RLO), the mass transfer can become dynamically unstable \citep{savonije1978}, triggering a common-envelope (CE) phase \citep{Paczynski1976, Ivanova2013}. During the CE phase, dynamical friction acting on the compact object (neutron star or black hole) leads to rapid orbital decay. If the hydrogen-rich envelope is successfully ejected before stellar coalescence occurs, the surviving binary consists of the compact remnant and the exposed helium core of the donor star. Following the subsequent SN explosion of the helium star, a BNS or NSBH system is formed, which will eventually inspiral due to gravitational-wave (GW) emission. Provided the post-SN orbit is sufficiently tight and eccentric, a subset of these systems will merge within a Hubble time, contributing to the population of compact binaries detected by the LVK detector network (see, e.g., \citealt{Mapelli2021, MandelBroekgaarden2021} for reviews of the IBE and other formation channels).

LRNe, believed to be associated with the ejection of CE, thus offers a unique observational window into probing the CE phase (first observational evidence from V1309 Sco, \citealt{tylenda2011}).
LRNe can therefore provide valuable insight into the formation of the compact binaries, and enhance our understanding of the CEE phase. Recently, using the systematic sample from the Census of the Local Universe (CLU) experiment \citep{De2020} of the Zwicky Transient Facility \citep[ZTF;][]{Bellm2019}, the local rate of LRNe was estimated to be $\rho_0=7.8^{+6.5}_{-3.7} \times 10^{-5} ~\mathrm{Mpc^{-3}}~\mathrm{yr^{-1}}$ in the magnitude range of $-16 \leq M_V \leq -11$ \citep{karambelkar2023}.

The rate of LRNe that results in the formation of a BNS or NSBH binary that merges in Hubble time should be related to the rate of CBCs estimated from LVK observations of such mergers. Comparing the rates of LRNe and BNS/NSBH mergers, therefore, allows us to place constraints on the fraction of LRNe that form compact binaries that merge within the age of the Universe. 

In this work, we put forward a model that relates the volumetric LRNe rates to the corresponding merger rates of BNS and NSBH binaries and compare this with the observed volumetric rates of BNS and NSBH mergers from LVK observations. As mentioned before, we do not consider BBH mergers here as their rates may have a non-negligible contribution from dynamical formation channels and therefore are not as clean as BNS and NSBH mergers whose contribution from dynamical formation scenarios is believed to be negligible~\citep[see, e.g.,][]{Ye2019,Clausen2012}. More specifically, our goal is to estimate the fraction of LRNe that are associated with the population of BNS and NSBHs detected by LVK. 

From here onward, for brevity, CBCs will refer to BNSs and NSBHs exclusively. The detailed methodology followed is described in Section~\ref{method}.
\begin{table*}[htb]
    \centering
    \captionsetup{justification=centering}
    \caption{Various delay time models employed and the corresponding free parameters of the models.}
    \renewcommand{\arraystretch}{1.8}
    \begin{tabular}{|c|c|c|c|}
        \hline
        \textbf{} & \textbf{Expression} & \textbf{Parameters} & \textbf{References} \\
        \hline
        Power Law & $P(t_d | t_d^{\text{min}}) = 1/t_d$ & $t_d > t_d^{\text{min}},\ t_d^{\text{min}} \sim 20\ \text{Myr}$ & \cite{piran1992,Belczynski2006}. \\
        \hline
        Exponential & $P(t_d | \tau) = \frac{1}{\tau} \exp\left(-\frac{t_d}{\tau}\right)$ & $\tau = 0.1\ \text{Gyr}$ & \cite{vitale2019} \\
        \hline
        Lognormal & $P(t_d | \tau, \sigma_t) = \frac{1}{\sqrt{2\pi}\sigma_t} \exp\left(-\frac{(\ln t_d - \ln \tau)^2}{2\sigma_t^2}\right)$ & $\tau = 2.9\ \text{Gyr}$ & \cite{wanderman2015} \\
        & & $\sigma_t = 0$ & \\
        \hline
    \end{tabular}
    \label{tab:distributions}
\end{table*}
\section{Method}
\label{method}

Our goal is to compare the observed rate of CBCs as inferred from LVK observations  $R_{\rm CBC}^{\rm LVK}(z=0)$ with the rate projected from LRNe observations $R_{\rm CBC}^{\rm LRNe} (z=0)$ with the aim of understanding what fraction of the LRNe lead to CBCs. To obtain the former, we assume that the intrinsic rate of LRNe %(denoted by $R_{\rm LRNe}$) 
at any redshift closely follows the star formation rate $\rho(z)$ at that redshift. It is convenient to rewrite:
\begin{equation}
    \rho(z)=\rho_0\,f(z)
\end{equation}
where $\rho_0$ is the SFR at $z=0$. Because the LRNe rate closely follows SFR, we can write down the {\it rate of formation of compact binaries} (CBF) from LRNe as:
\begin{equation}\label{Eq:lrne_ansatz}
    \rho_{\rm CBF}^{\rm LRNe}(z)=\rho_0^{\rm LRNe}\,f(z)
\end{equation}
where $\rho_0^{\rm LRNe}$ is provided in \cite {karambelkar2023} and $f(z)$ denotes the redshift evolution function of this rate following a SFR. Here, we assume one to one correspondence between the compact binaries and the LRNe, that is, one LRNe event will be associated to a single compact binary formation. The compact binaries (CBs) formed by this mechanism will merge under GW radiation~\citep{Peters1963}, and the time taken for a CB to merge is called the delay time $t_d$. The delay time crucially depends on the orbital parameters of the CB at the time of formation. A tighter binary will merge faster than a wider binary, and the same results for eccentric orbits compared to circular ones~\citep{Peters1964}. As the delay time distribution of CBs formed via isolated binary evolution is unknown, we employ different models and assess the impact of this assumption on our estimates. 

The projected CBC merger rate from the LRNe rate can be obtained by convolving the SFR with the delay time distribution and integrating over the delay time. The resulting CBC merge rate can be written as:
\begin{equation} 
R_{\rm CBC}^{\rm LRNe}(z)=\int_{t_d^{min}}^{\infty}\frac{\rho_{\rm CBF}^{\rm LRNe}(z_f(z,t_d))}{1+z_f(z,t_d)}\,P(t_d)dt_d,\label{eq:Rcbc}
\end{equation}
where $z_f$ denotes the redshift at formation of the binary and $P(t_d)$ denotes the delay time distribution we employ which has a $t_d^{\rm{min}}$, the minimum delay time assumed (see Table~\ref{tab:distributions} for details). Because redshift at merger depends on $z_f$ and the delay time, one can invert the relation to obtain $z_f(z,t_d)$, which is what appears in the above expression. From Eq.~(\ref{eq:Rcbc}), we can straightforwardly obtain $\rho_{\rm{CBC}}^{\rm LRNe}(z=0)$, the projected merger rate of CBCs at the present epoch from the LRNe rates. 

We can now compare the LRNe-based CBC rate estimate $\rho_{\rm CBC}^{\rm LRNe}(z=0)$ with $R_{\rm CBC}^{\rm LVK}$, the inferred CBC merger rate from LVK observations. We compute the ratio of these two rates, defined as:
\begin{equation}
    f^{\rm LVK}_{\rm LRNe}=\frac{R_{\rm CBC}^{\rm LVK}(z=0)}{R_{\rm CBC}^{\rm LRNe}(z=0)},\label{eq:frac}
\end{equation}
where the CBC population from LVK observation contains BNS and NSBH as reported in Table~2 of \cite{GWTC-4Rates}. In particular, we use the posterior samples pertaining to the BNS/NSBH \textsc{FullPop-4.0}  model \citep{GWTC-4Rates} made available with the paper \citep{GWTC-4zenodo}.

We present two classes of results. In the first, we make use of the plot of the LRNe rate as a function of their absolute magnitudes given in Figure~12 of \cite{karambelkar2023}. This allows us to compute the fraction $f^{\rm LVK}_{\rm LRNe}$ as a function of the absolute magnitude of the LRNe, helping us understand the dependence of this fraction on their magnitudes. In the second set of results, we obtain the posterior distribution of this fraction, using the aggregated rate of LRNe across all magnitudes. 

\begin{figure}[H]
\includegraphics[width=\linewidth]{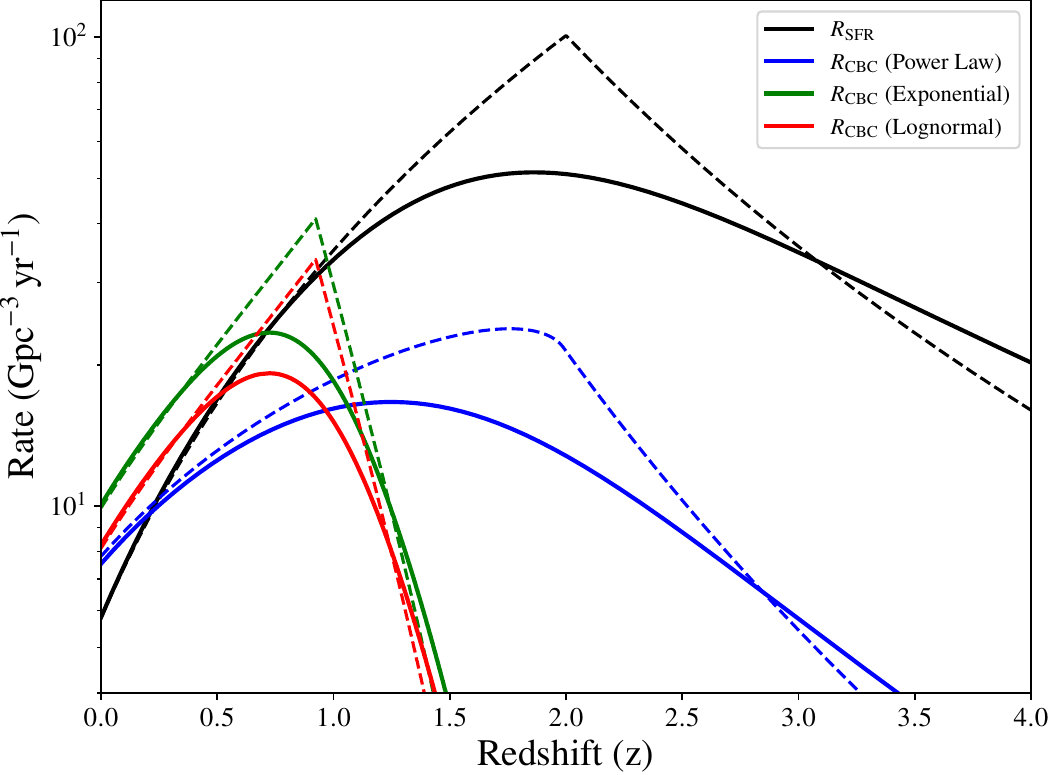}
\caption{Star formation rate along with the merger rate. Solid lines represents \cite{MadauDickinson2014}, whereas the dashed lines represents the work of \cite{wanderman2015}. The SFR is given in the color black, the blue, green and red lines represents the CBC rate with the power law, lognormal and exponential delay distributions, respectively.}
\label{fig:SFR/SGRB dist}
\end{figure}

For this work, we have used two star formation rates (SFR). The first, based on \cite{MadauDickinson2014}, 
combines data from far-UV, mid-IR, and far-IR to construct a functional representation of the cosmic SFR, though limited to redshifts $z<8$:
\begin{equation} \label{eq:SFR}
    \rho_{\rm SFR}=\rho_0\frac{(1+z)^{2.7}}{1+[(1+z)/2.9]^{5.6}}~\mathrm{M_{\odot}}~\mathrm{yr^{-1}} ~\mathrm{Mpc^{-3}}
\end{equation}
with $\rho_0 = 0.015$. 
The second one is based on \cite{Cucciati2012}, which uses the VIMOS-VLT Deep Survey, a single deep galaxy redshift survey up to redshift $\mathrm{z}\sim 4.5$:
\begin{equation} \label{eq:SFR2}
            \rho_{\rm SFR} =
            \begin{cases}
            c_1(1+z)^{2.6}, & \text{if }~~~~~ z \leq {2} \\
            c_2(1+z)^{-3.6}, & \text{if }~~~~~ z \geq {2} \\
            \end{cases}
            ~~~\mathrm{M_{\odot}}~\mathrm{yr^{-1}} ~\mathrm{Mpc^{-3}}
        \end{equation}
This SFR is used for the verification of the method applied in this work, where we compare the CBC merger rate with  \cite{wanderman2015}, which uses the similar approach for short GRB rate estimation using the peak fluxes and redshifts of {\textit{BATSE}}, \textit{Swift} and \textit{Fermi} (Figure \ref{fig:SFR/SGRB dist}).

We use three delay time distribution $P(t_d)$ models: power law, lognormal and exponential (see Table~\ref{tab:distributions} for details). Although the delay time distribution models adopted in this work are well established in studies of compact binary mergers, it is important to note that natal kicks associated with the common envelope (CE) ejection process may alter the resulting merger timescales \citep{giacobbo2018}. The SFR and the CBC merger rates for these models are depicted in Figure~\ref{fig:SFR/SGRB dist}. We observe broad consistency between CBC rates evaluated using different delay time distribution models, particularly at low redshifts, where the fraction of LRNe associated with CBCs is examined.

The local event rate at $M_V \leq -11$, $\rho^{\rm LRNe}_0=7.8^{+6.5}_{-3.7} \times 10^{-5} ~\mathrm{Mpc^{-3}}~\mathrm{yr^{-1}}$, is based on the LRNe rate estimated by \cite{karambelkar2023}. The rate scales as $\frac{dN}{dL} \propto L^{-2.5}$, which can be converted for the magnitude range of $-16 \leq M_V \leq -11$, using the relation $L \propto 10^{-0.4M}$. The LRNe redshift dependent rate is then assumed to be $\rho_{\rm SFR}/\rho_0 \times \rho_0^{\rm LRNe}$ (cf. Eqs.~\ref{Eq:lrne_ansatz}, \ref{eq:SFR}).

\begin{figure*}[htb]
\includegraphics[width=\linewidth]{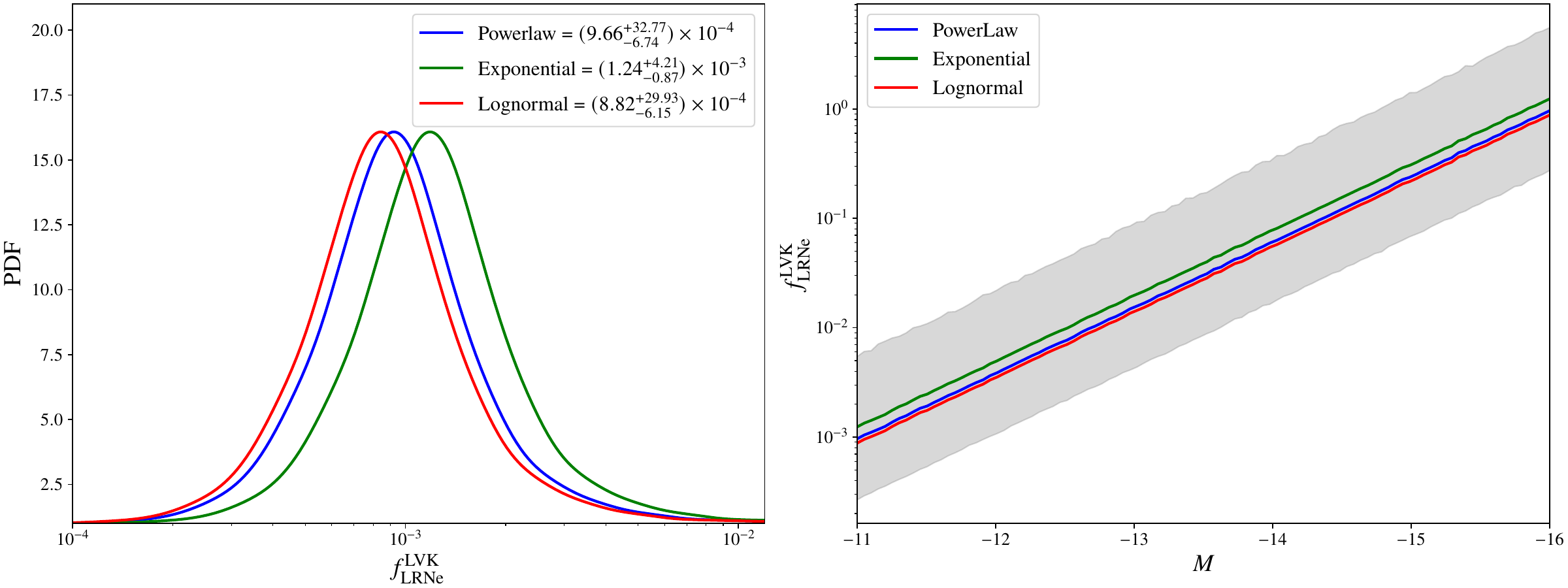}
\caption{{\bf Left:} Posteriors of the ratio of the sum of BNS and NSBH mergers observed by the LVK collaboration with the CBC rates calculated from LRNe rates (see Eq. \ref{eq:frac}) for the different delay distributions. The LRNe are assumed to follow the SFR estimated by \cite{MadauDickinson2014}. {\bf Right:} Fraction of LRNe brighter than the absolute magnitude $M_V=-11$ resulting in BNS/NSBH mergers. The trend shows the fraction of events brighter than the given absolute magnitude, with the shaded region showing the $90\%$ confidence interval.}
\label{fig:pl_R_(bns+nsbh)/R_tot}
\end{figure*}

\section{Results and Discussion}

First, we consider the LRNe as a single population and ask what fraction of them can be associated with the LVK-detected CBC population. The posterior distribution of this fraction (again, see Eq.~(\ref{eq:frac})) is shown in the left panel of Figure~\ref{fig:pl_R_(bns+nsbh)/R_tot} for different delay time distributions. To construct this distribution, from the LVK point, we consider the BNS and NSBH rates given in Table~2 of~\citep{GWTC-4Rates} corresponding to the {\tt FULLPOP-4.0}, the sum of which is $R^{\rm LVK}_{\rm CBC}$ based on our definition. The rate posterior of this sum is acquired by adding, sample by sample, the rate posteriors provided in \cite{GWTC-4zenodo}. The LRNe rate samples are constructed by drawing from a skewed gaussian distribution with mean and standard deviation provided by the rate estimates and error bars in \cite{karambelkar2023}. The fraction is then simply the ratio distribution of the LVK's CBC rate and the LRNe rate. We find that the typical median value of this fraction, regardless of the delay time distribution model, is $ {\cal O}(10^{-3})$. 

The small value of this fraction has profound implications. It suggests that the vast majority of LRNe do not result in CBCs that merge in Hubble time. Considering stellar mergers and compact binary formation as two leading candidates for the end products of LRNe, this could imply that stellar mergers dominate significantly over the CB formation, although CBs that fail to merge in Hubble time cannot be ruled out as the dominant end product of LRNe, exclusively from the value of this fraction.

Another intriguing possibility is that the less luminous LRNe correspond to stellar mergers, whereas the most luminous ones correspond to CBCs that merge in Hubble time. Figure~12 of \cite{karambelkar2023} presents the volumetric rate of LRNe brighter than a given peak absolute magnitude: $\rho_0^{\rm LRNe}\geq M_{\rm peak}$ versus $M_{\rm peak}$ in our notation. Based on the prescription outlined in Section~\ref{method}, we can convert this into a CBC merger rate predicted from LRNe, $R_{\rm CBC}^{\rm LRNe} \geq M_{\rm peak}$, accounting for the error bars quoted in \cite{karambelkar2023}. Dividing this by the LVK rates for CBCs, we obtain $f^{\rm LRNe}_{\rm LVK}$, defined in Eq.~(\ref{eq:frac}), as a function of  $M_{\rm peak}$. 

The result is shown in the right panel of Figure~\ref{fig:pl_R_(bns+nsbh)/R_tot}. While this is based on the \cite{MadauDickinson2014} SFR model, we find that using the \cite{Cucciati2012} model does not change the results by more than $5\%$. It is evident from the plot that the LVK CBC rate contributes negligibly to the rate of less luminous LRNe. On the other hand, the subpopulation of more luminous LRNe, say with $M_{\rm peak}\leq -14$, can constitute between $10-100\%$ of the LVK-detected CBC population. It is interesting to note that the rates of the brightest LRNe are consistent with the rates of CBCs, implying that, if all CBCs are indeed associated with LRNe, then they're likely associated only with the brightest of them. Conversely, if LRNe do produce merged CBCs as end products, then only the brightest ones do, while the majority of the rest -- the dimmer LRNe -- are unrelated to merged CBCs, possibly because they pertain to stellar mergers. 

Finally, we wish to point out that even if a considerable fraction of LRNe produce tight stellar binaries, the natal kicks arising from mass ejection from asymmetric explosions driven by neutrino emission during the supernova explosion can still prevent the formation of compact binaries\citep{Brandt1995, Kalogera1998, Renzo2019}. Owing to the complexities associated with supernova kicks, it is not straightforward to estimate this fraction. A better understanding of the kick-velocity distribution in compact binaries in the future will help clarify the role of this effect.

In our analysis, we have assumed that all binaries formed have one associated LRN. Although possible, it is highly unlikely for the binary to have two unstable mass transfer events which would not result in a merger.
We have also assumed that all CBCs (BNSs/NSBHs) are formed via IBE. However, these assumptions may overestimate the true fraction of LRNe, as not all IBE binaries undergo interactions that produce observable luminous transients, and a hitherto unknown fraction of CBCs (esp. NSBHs) may potentially be formed in dense stellar environments without requiring a CE phase. Specifically, LRNe are typically associated with CE ejection or stellar mergers—events that occur only in a subset of IBE scenarios. Systems that evolve through stable mass transfer or those with compact, high-mass primaries may not generate an LRNe, either due to the absence of envelope ejection or because the ejection is not radiatively efficient. As such, our results should be interpreted as upper limits on the fraction of LRNe that produce CBCs as end products. 

It is important to note that the calculations have been carried out for mergers occurring in the local universe, i.e., at $z=0$, due to the lack of redshift-dependent estimates for BNS/NSBH merger rates from the LVK. However, with the advent of third-generation GW detectors, it may become possible to explore the redshift dependence of these rates.

\section{Conclusion}
We have utilised GW observations of BNS and NSBH mergers as a new tool to understand the intriguing class of gap transients known as LRNe. Despite the established association of LRNe with CE ejection in a stellar binary evolution, the end product of LRNe is an open question in transient astronomy. Leading contenders for possible end products include stellar mergers and the formation of tighter stellar binaries, capable of producing compact binaries -- a fraction of which will merge within Hubble time. We propose a simple model to connect the compact binary merger rate estimated from GW observations to the LRNe rate, if the LRNe give rise to a compact binary. Using our method, we find only ${\cal O}(10^{-3})$ of the LRNe are associated with CBC mergers seen in GWs. This could imply that a significantly large fraction of LRNe observed to date could correspond to stellar mergers. We also find that the brighter the LRNe, the more likely they are to be associated with CBC mergers. Therefore, only a small subpopulation with high luminosity may be associated with compact binary formation. However, to make this association more certain, we require further investigation. 

The novelty of our result stems from the use of GW observations, which provide a completely new handle on this intriguing class of transients. As both LRNe and GW observations mature further, tighter and more interesting constraints can be placed on this fraction. This also signifies the growing importance of GW observations, which are answering open questions in transient astronomy in previously unforeseen ways.

The current sample of LRNe is limited and largely confined to the local universe. However, future observations from upcoming facilities, such as the Vera Rubin Observatory, are expected to significantly expand the LRN population. This will offer deeper insights into their redshift distribution, luminosity characteristics, and progenitor system properties. In tandem with future GW observations of BNSs/NSBHs at higher redshifts, our understanding of the connection between LRNe and CBCs will become even clearer and more complete. 

\section{Acknowldegments}
We thank Saleem Muhammed for his review and feedback on the manuscript. We thank Mansi M. Kasliwal for useful discussions. SJK acknowledges support from ANRF/SERB Grants SRG/2023/000419 and MTR/2023/000086. KM, LR and KGA acknowledge the support from the BRICS grant DST/ICD/BRICS/Call-5/CoNMuTraMO/2023 (G) funded by the Department of Science and Technology (DST), India. Dimple acknowledges support from the STFC grant No. ST/Y002253/1.

This research has made use of data or software obtained from the Gravitational Wave Open Science Center (gwosc.org), a service of the LIGO Scientific Collaboration, the Virgo Collaboration, and KAGRA. This material is based upon work supported by NSF's LIGO Laboratory which is a major facility fully funded by the National Science Foundation, as well as the Science and Technology Facilities Council (STFC) of the United Kingdom, the Max-Planck-Society (MPS), and the State of Niedersachsen/Germany for support of the construction of Advanced LIGO and construction and operation of the GEO600 detector. Additional support for Advanced LIGO was provided by the Australian Research Council. Virgo is funded, through the European Gravitational Observatory (EGO), by the French Centre National de Recherche Scientifique (CNRS), the Italian Istituto Nazionale di Fisica Nucleare (INFN) and the Dutch Nikhef, with contributions by institutions from Belgium, Germany, Greece, Hungary, Ireland, Japan, Monaco, Poland, Portugal, and Spain. KAGRA is supported by the Ministry of Education, Culture, Sports, Science and Technology (MEXT), Japan Society for the Promotion of Science (JSPS) in Japan; National Research Foundation (NRF) and Ministry of Science and ICT (MSIT) in Korea; Academia Sinica (AS) and National Science and Technology Council (NSTC) in Taiwan.

\bibliography{references}

@INPROCEEDINGS{Paczynski1976,
  author    = {Paczynski, B.},
  title     = {Evolutionary processes in close binary systems},
  booktitle = {Structure and Evolution of Close Binary Systems},
  editor    = {Eggleton, Peter and Mitton, Simon and Whelan, John},
  series    = {IAU Symposium},
  volume    = {73},
pages= "75",
  year      = {1976},
  address   = {Cambridge, England},
  month     = jul,
  publisher = {Reidel},
  adsurl    = {https://ui.adsabs.harvard.edu/abs/1976IAUS...73.....E},
}

@ARTICLE{Ivanova2013,
       author = {{Ivanova}, N. and {Justham}, S. and {Chen}, X. and {De Marco}, O. and {Fryer}, C.~L. and {Gaburov}, E. and {Ge}, H. and {Glebbeek}, E. and {Han}, Z. and {Li}, X. -D. and {Lu}, G. and {Marsh}, T. and {Podsiadlowski}, P. and {Potter}, A. and {Soker}, N. and {Taam}, R. and {Tauris}, T.~M. and {van den Heuvel}, E.~P.~J. and {Webbink}, R.~F.},
        title = "{Common envelope evolution: where we stand and how we can move forward}",
      journal = {\aapr},
         year = 2013,
        month = feb,
       volume = {21},
          eid = {59},
        pages = {59},
          doi = {10.1007/s00159-013-0059-2},
archivePrefix = {arXiv},
       eprint = {1209.4302},
 primaryClass = {astro-ph.HE},
       adsurl = {https://ui.adsabs.harvard.edu/abs/2013A&ARv..21...59I},
}

@article{Olejak2021,
    author = "Olejak, Aleksandra and Belczynski, Krzysztof and Ivanova, Natalia",
    title = "{Impact of common envelope development criteria on the formation of LIGO/Virgo sources}",
    eprint = "2102.05649",
    archivePrefix = "arXiv",
    primaryClass = "astro-ph.HE",
    doi = "10.1051/0004-6361/202140520",
    journal = "Astron. Astrophys.",
    volume = "651",
    pages = "A100",
    year = "2021"
}

@article{Postnov2014,
    author = "Postnov, Konstantin A. and Yungelson, Lev R.",
    title = "{The Evolution of Compact Binary Star Systems}",
    eprint = "1403.4754",
    archivePrefix = "arXiv",
    primaryClass = "astro-ph.HE",
    doi = "10.12942/lrr-2014-3",
    journal = "Living Rev. Rel.",
    volume = "17",
    pages = "3",
    year = "2014"
}

@ARTICLE{Politano2010,
       author = {{Politano}, Michael and {van der Sluys}, Marc and {Taam}, Ronald E. and {Willems}, Bart},
        title = "{Population Synthesis of Common Envelope Mergers. I. Giant Stars with Stellar or Substellar Companions}",
      journal = {\apj},
         year = 2010,
        month = sep,
       volume = {720},
       number = {2},
        pages = {1752-1766},
          doi = {10.1088/0004-637X/720/2/1752},
archivePrefix = {arXiv},
       eprint = {1007.4545},
 primaryClass = {astro-ph.SR},
       adsurl = {https://ui.adsabs.harvard.edu/abs/2010ApJ...720.1752P}
}

@article{Howitt2019,
    author = "Howitt, George and Stevenson, Simon and Vigna-G{\'o}mez, Alejandro and Justham, Stephen and Ivanova, Natasha and Woods, Tyrone E. and Neijssel, Coenraad J. and Mandel, Ilya",
    title = "{Luminous Red Novae: population models and future prospects}",
    eprint = "1912.07771",
    archivePrefix = "arXiv",
    primaryClass = "astro-ph.HE",
    doi = "10.1093/mnras/stz3542",
    journal = "Mon. Not. Roy. Astron. Soc.",
    volume = "492",
    number = "3",
    pages = "3229--3240",
    year = "2020"
}

@article{GWTC-4,
    author = "Abac, A. G. and others",
    collaboration = "LIGO Scientific, VIRGO, KAGRA",
    title = "{GWTC-4.0: Updating the Gravitational-Wave Transient Catalog with Observations from the First Part of the Fourth LIGO-Virgo-KAGRA Observing Run}",
    eprint = "2508.18082",
    archivePrefix = "arXiv",
    primaryClass = "gr-qc",
    reportNumber = "LIGO-P2400386",
    month = "8",
    year = "2025"
}

@article{GWTC-4Rates,
    author = "Abac, A. G. and others",
    collaboration = "LIGO Scientific, VIRGO, KAGRA",
    title = "{GWTC-4.0: Population Properties of Merging Compact Binaries}",
    eprint = "2508.18083",
    archivePrefix = "arXiv",
    primaryClass = "astro-ph.HE",
    reportNumber = "LIGO-P2400004",
    month = "8",
    year = "2025"
}

@article{Belczynski2017,
    author = "Belczynski, K. and others",
    title = "{The origin of the first neutron star {\textendash} neutron star merger}",
    eprint = "1712.00632",
    archivePrefix = "arXiv",
    primaryClass = "astro-ph.HE",
    doi = "10.1051/0004-6361/201732428",
    journal = "Astron. Astrophys.",
    volume = "615",
    pages = "A91",
    year = "2018"
}

@article{Ye2019,
    author = "Ye, Claire S. and Fong, Wen-fai and Kremer, Kyle and Rodriguez, Carl L. and Chatterjee, Sourav and Fragione, Giacomo and Rasio, Frederic A.",
    title = "{On the Rate of Neutron Star Binary Mergers from Globular Clusters}",
    eprint = "1910.10740",
    archivePrefix = "arXiv",
    primaryClass = "astro-ph.HE",
    doi = "10.3847/2041-8213/ab5dc5",
    journal = "Astrophys. J. Lett.",
    volume = "888",
    number = "1",
    pages = "L10",
    year = "2020"
}

@article{Dominik2012,
    author = "Dominik, Michal and Belczynski, Krzysztof and Fryer, Christopher and Holz, Daniel and Berti, Emanuele and Bulik, Tomasz and Mandel, Ilya and O'Shaughnessy, Richard",
    title = "{Double Compact Objects I: The Significance of the Common Envelope on Merger Rates}",
    eprint = "1202.4901",
    archivePrefix = "arXiv",
    primaryClass = "astro-ph.HE",
    doi = "10.1088/0004-637X/759/1/52",
    journal = "Astrophys. J.",
    volume = "759",
    pages = "52",
    year = "2012"
}

@article{GW150914,
    author = "Abbott, B. P. and others",
    collaboration = "LIGO Scientific, Virgo",
    title = "{Observation of Gravitational Waves from a Binary Black Hole Merger}",
    eprint = "1602.03837",
    archivePrefix = "arXiv",
    primaryClass = "gr-qc",
    reportNumber = "LIGO-P150914",
    doi = "10.1103/PhysRevLett.116.061102",
    journal = "Phys. Rev. Lett.",
    volume = "116",
    number = "6",
    pages = "061102",
    year = "2016"
}

@article{GW170817,
    author = "Abbott, B. P. and others",
    collaboration = "LIGO Scientific, Virgo",
    title = "{GW170817: Observation of Gravitational Waves from a Binary Neutron Star Inspiral}",
    eprint = "1710.05832",
    archivePrefix = "arXiv",
    primaryClass = "gr-qc",
    reportNumber = "LIGO-P170817",
    doi = "10.1103/PhysRevLett.119.161101",
    journal = "Phys. Rev. Lett.",
    volume = "119",
    number = "16",
    pages = "161101",
    year = "2017"
}

@article{GW190425,
    author = "Abbott, B. P. and others",
    collaboration = "LIGO Scientific, Virgo",
    title = "{GW190425: Observation of a Compact Binary Coalescence with Total Mass $\sim 3.4 M_{\odot}$}",
    eprint = "2001.01761",
    archivePrefix = "arXiv",
    primaryClass = "astro-ph.HE",
    reportNumber = "LIGO-P190425",
    doi = "10.3847/2041-8213/ab75f5",
    journal = "Astrophys. J. Lett.",
    volume = "892",
    number = "1",
    pages = "L3",
    year = "2020"
}

@article{LVK-NSBH,
    author = "Abbott, R. and others",
    collaboration = "LIGO Scientific, KAGRA, VIRGO",
    title = "{Observation of Gravitational Waves from Two Neutron Star{\textendash}Black Hole Coalescences}",
    eprint = "2106.15163",
    archivePrefix = "arXiv",
    primaryClass = "astro-ph.HE",
    reportNumber = "LIGO Document P2000357",
    doi = "10.3847/2041-8213/ac082e",
    journal = "Astrophys. J. Lett.",
    volume = "915",
    number = "1",
    pages = "L5",
    year = "2021"
}

@article{kasliwal2011,
doi = {10.1088/0004-637X/735/2/94},
url = {https://doi.org/10.1088/0004-637X/735/2/94},
year = {2011},
month = {jun},
publisher = {The American Astronomical Society},
volume = {735},
number = {2},
pages = {94},
author = {Kasliwal, M. M. and Cenko, S. B. and Kulkarni, S. R. and Ofek, E. O. and Quimby, R. and Rau, A.},
title = {DISCOVERY OF A NEW PHOTOMETRIC SUB-CLASS OF FAINT AND FAST CLASSICAL NOVAE},
journal = {The Astrophysical Journal},
}

@article{aligo,
    author = "Aasi, J. and others",
    collaboration = "LIGO Scientific",
    title = "{Advanced LIGO}",
    eprint = "1411.4547",
    archivePrefix = "arXiv",
    primaryClass = "gr-qc",
    doi = "10.1088/0264-9381/32/7/074001",
    journal = "Class. Quant. Grav.",
    volume = "32",
    pages = "074001",
    year = "2015"
}

@article{avirgo,
    author = "Acernese, F. and others",
    collaboration = "VIRGO",
    title = "{Advanced Virgo: a second-generation interferometric gravitational wave detector}",
    eprint = "1408.3978",
    archivePrefix = "arXiv",
    primaryClass = "gr-qc",
    doi = "10.1088/0264-9381/32/2/024001",
    journal = "Class. Quant. Grav.",
    volume = "32",
    number = "2",
    pages = "024001",
    year = "2015"
}

@article{Clausen2012,
    author = "Clausen, Drew and Sigurdsson, Steinn and Chernoff, David F.",
    title = "{Black Hole-Neutron Star Mergers in Globular Clusters}",
    eprint = "1210.8153",
    archivePrefix = "arXiv",
    primaryClass = "astro-ph.HE",
    doi = "10.1093/mnras/sts295",
    journal = "Mon. Not. Roy. Astron. Soc.",
    volume = "428",
    pages = "3618",
    year = "2013"
}

@article{Peters1963,
     author        = {Peters, P.C. and Mathews, J.},
     title         = {Gravitational Radiation from Point Masses in a Keplerian Orbit},
     journal       = {Phys. Rev.},
     volume        = {131},
     pages         = {435--440},
     year          = {1963},
     ReferencedIn  = {2002-3blanchet}
}

@article{Peters1964,
     author        = {Peters, P.C.},
     title         = {Gravitational Radiation and the Motion of Two Point Masses},
     journal       = {Phys. Rev.},
     volume        = {136},
     pages         = {B1224--B1232},
     year          = {1964},
     ReferencedIn  = {2002-3blanchet}
}

@article{KAGRA,
    author = "Akutsu, T. and others",
    collaboration = "KAGRA",
    title = "{Overview of KAGRA: Detector design and construction history}",
    eprint = "2005.05574",
    archivePrefix = "arXiv",
    primaryClass = "physics.ins-det",
    doi = "10.1093/ptep/ptaa125",
    journal = "PTEP",
    volume = "2021",
    number = "5",
    pages = "05A101",
    year = "2021"
}

@article{pastorello2019,
       author = {{Pastorello}, A. and {Mason}, E. and {Taubenberger}, S. and {Fraser}, M. and {Cortini}, G. and {Tomasella}, L. and {Botticella}, M.~T. and {Elias-Rosa}, N. and {Kotak}, R. and {Smartt}, S.~J. and {Benetti}, S. and {Cappellaro}, E. and {Turatto}, M. and {Tartaglia}, L. and {Djorgovski}, S.~G. and {Drake}, A.~J. and {Berton}, M. and {Briganti}, F. and {Brimacombe}, J. and {Bufano}, F. and {Cai}, Y.-Z. and {Chen}, S. and {Christensen}, E.~J. and {Ciabattari}, F. and {Congiu}, E. and {Dimai}, A. and {Inserra}, C. and {Kankare}, E. and {Magill}, L. and {Maguire}, K. and {Martinelli}, F. and {Morales-Garoffolo}, A. and {Ochner}, P. and {Pignata}, G. and {Reguitti}, A. and {Sollerman}, J. and {Spiro}, S. and {Terreran}, G. and {Wright}, D.~E.},
        title = "{Luminous red novae: Stellar mergers or giant eruptions?}",
      journal = {\aap},
         year = 2019,
        month = oct,
       volume = {630},
          eid = {A75},
        pages = {A75},
          doi = {10.1051/0004-6361/201935999},
archivePrefix = {arXiv},
       eprint = {1906.00812},
 primaryClass = {astro-ph.SR},
       adsurl = {https://ui.adsabs.harvard.edu/abs/2019A&A...630A..75P},
}

@article{tylenda2005,
  title={Evolution of V838 Monocerotis during and after the 2002 eruption},
  author={Tylenda, R},
  journal={Astronomy \& Astrophysics},
  volume={436},
  number={3},
  pages={1009--1020},
  year={2005},
  publisher={EDP Sciences}
}

@article{tylenda2011,
  title={V1309 Scorpii: merger of a contact binary},
  author={Tylenda, R and Hajduk, M and Kami{\'n}ski, T and Udalski, A and Soszy{\'n}ski, I and Szyma{\'n}ski, MK and Kubiak, M and Pietrzy{\'n}ski, G and Poleski, R and Ulaczyk, K and others},
  journal={Astronomy \& Astrophysics},
  volume={528},
  pages={A114},
  year={2011},
  publisher={EDP Sciences}
}

@ARTICLE{MadauDickinson2014,
       author = {{Madau}, Piero and {Dickinson}, Mark},
        title = "{Cosmic Star-Formation History}",
      journal = {\araa},
         year = 2014,
        month = aug,
       volume = {52},
        pages = {415-486},
          doi = {10.1146/annurev-astro-081811-125615},
archivePrefix = {arXiv},
       eprint = {1403.0007},
 primaryClass = {astro-ph.CO},
       adsurl = {https://ui.adsabs.harvard.edu/abs/2014ARA&A..52..415M},
}

@article{piran1992,
  title={The implications of the Compton (GRO) observations for cosmological gamma-ray bursts},
  author={Piran, Tsvi},
  journal={Astrophysical Journal, Part 2-Letters (ISSN 0004-637X), vol. 389, April 20, 1992, p. L45-L48.},
  volume={389},
  pages={L45--L48},
  year={1992}
}

@article{vitale2019,
  title={Measuring the star formation rate with gravitational waves from binary black holes},
  author={Vitale, Salvatore and Farr, Will M and Ng, Ken KY and Rodriguez, Carl L},
  journal={The Astrophysical Journal Letters},
  volume={886},
  number={1},
  pages={L1},
  year={2019},
  publisher={IOP Publishing}
}

@article{wanderman2015,
  title={The rate, luminosity function and time delay of non-Collapsar short GRBs},
  author={Wanderman, David and Piran, Tsvi},
  journal={Monthly Notices of the Royal Astronomical Society},
  volume={448},
  number={4},
  pages={3026--3037},
  year={2015},
  publisher={Oxford University Press}
}

@article{Belczynski2006,
doi = {10.1086/505169},
url = {https://doi.org/10.1086/505169},
year = {2006},
month = {sep},
publisher = {},
volume = {648},
number = {2},
pages = {1110},
author = {Belczynski, Krzysztof and Perna, Rosalba and Bulik, Tomasz and Kalogera, Vassiliki and Ivanova, Natalia and Lamb, Donald Q.},
title = {A Study of Compact Object Mergers as Short Gamma-Ray Burst Progenitors},
journal = {The Astrophysical Journal},
}

@article{karambelkar2023,
doi = {10.3847/1538-4357/acc2b9},
url = {https://doi.org/10.3847/1538-4357/acc2b9},
year = {2023},
month = {may},
publisher = {The American Astronomical Society},
volume = {948},
number = {2},
pages = {137},
author = {Karambelkar, Viraj R. and Kasliwal, Mansi M. and Blagorodnova, Nadejda and Sollerman, Jesper and Aloisi, Robert and Anand, Shreya G. and Andreoni, Igor and Brink, Thomas G. and Bruch, Rachel and Cook, David and Das, Kaustav Kashyap and De, Kishalay and Drake, Andrew and Filippenko, Alexei V. and Fremling, Christoffer and Helou, George and Ho, Anna and Jencson, Jacob and Jones, David and Laher, Russ R. and Masci, Frank J. and Patra, Kishore C. and Purdum, Josiah and Reedy, Alexander and Sit, Tawny and Sharma, Yashvi and Tzanidakis, Anastasios and van der Walt, Stéfan J. and Yao, Yuhan and Zhang, Chaoran},
title = {Volumetric Rates of Luminous Red Novae and Intermediate-luminosity Red Transients with the Zwicky Transient Facility},
journal = {The Astrophysical Journal},
}

@article{De2020,
doi = {10.3847/1538-4357/abb45c},
url = {https://doi.org/10.3847/1538-4357/abb45c},
year = {2020},
month = {dec},
publisher = {The American Astronomical Society},
volume = {905},
number = {1},
pages = {58},
author = {De, Kishalay and Kasliwal, Mansi M. and Tzanidakis, Anastasios and Fremling, U. Christoffer and Adams, Scott and Aloisi, Robert and Andreoni, Igor and Bagdasaryan, Ashot and Bellm, Eric C. and Bildsten, Lars and Cannella, Christopher and Cook, David O. and Delacroix, Alexandre and Drake, Andrew and Duev, Dmitry and Dugas, Alison and Frederick, Sara and Gal-Yam, Avishay and Goldstein, Daniel and Golkhou, V. Zach and Graham, Matthew J. and Hale, David and Hankins, Matthew and Helou, George and Ho, Anna Y. Q. and Irani, Ido and Jencson, Jacob E. and Kaplan, David L. and Kaye, Stephen and Kulkarni, S. R. and Kupfer, Thomas and Laher, Russ R. and Leadbeater, Robin and Lunnan, Ragnhild and Masci, Frank J. and Miller, Adam A. and Neill, James D. and Ofek, Eran O. and Perley, Daniel A. and Polin, Abigail and Prince, Thomas A. and Quataert, Eliot and Reiley, Dan and Riddle, Reed L. and Rusholme, Ben and Sharma, Yashvi and Shupe, David L. and Sollerman, Jesper and Tartaglia, Leonardo and Walters, Richard and Yan, Lin and Yao, Yuhan},
title = {The Zwicky Transient Facility Census of the Local Universe. I. Systematic Search for Calcium-rich Gap Transients Reveals Three Related Spectroscopic Subclasses},
journal = {The Astrophysical Journal},
}

@article{Bellm2019,
doi = {10.1088/1538-3873/ab0c2a},
url = {https://doi.org/10.1088/1538-3873/ab0c2a},
year = {2019},
month = {apr},
publisher = {The Astronomical Society of the Pacific},
volume = {131},
number = {1000},
pages = {068003},
author = {Bellm, Eric C. and Kulkarni, Shrinivas R. and Barlow, Tom and Feindt, Ulrich and Graham, Matthew J. and Goobar, Ariel and Kupfer, Thomas and Ngeow, Chow-Choong and Nugent, Peter and Ofek, Eran and Prince, Thomas A. and Riddle, Reed and Walters, Richard and Ye, Quan-Zhi},
title = {The Zwicky Transient Facility: Surveys and Scheduler},
journal = {Publications of the Astronomical Society of the Pacific},
}

@ARTICLE{Cucciati2012,
       author = {{Cucciati}, O. and {Tresse}, L. and {Ilbert}, O. and {Le F{\`e}vre}, O. and {Garilli}, B. and {Le Brun}, V. and {Cassata}, P. and {Franzetti}, P. and {Maccagni}, D. and {Scodeggio}, M. and {Zucca}, E. and {Zamorani}, G. and {Bardelli}, S. and {Bolzonella}, M. and {Bielby}, R.~M. and {McCracken}, H.~J. and {Zanichelli}, A. and {Vergani}, D.},
        title = "{The star formation rate density and dust attenuation evolution over 12 Gyr with the VVDS surveys}",
      journal = {\aap},
         year = 2012,
        month = mar,
       volume = {539},
          eid = {A31},
        pages = {A31},
          doi = {10.1051/0004-6361/201118010},
archivePrefix = {arXiv},
       eprint = {1109.1005},
 primaryClass = {astro-ph.CO},
       adsurl = {https://ui.adsabs.harvard.edu/abs/2012A&A...539A..31C},
}

@ARTICLE{FGMC,
       author = {{Farr}, Will M. and {Gair}, Jonathan R. and {Mandel}, Ilya and {Cutler}, Curt},
        title = "{Counting and confusion: Bayesian rate estimation with multiple populations}",
      journal = {\prd},
         year = 2015,
        month = jan,
       volume = {91},
       number = {2},
          eid = {023005},
        pages = {023005},
          doi = {10.1103/PhysRevD.91.023005},
archivePrefix = {arXiv},
       eprint = {1302.5341},
 primaryClass = {astro-ph.IM},
       adsurl = {https://ui.adsabs.harvard.edu/abs/2015PhRvD..91b3005F},
}

@article{Kapadia2019,
    author = "Kapadia, Shasvath J. and others",
    title = "{A self-consistent method to estimate the rate of compact binary coalescences with a Poisson mixture model}",
    eprint = "1903.06881",
    archivePrefix = "arXiv",
    primaryClass = "astro-ph.HE",
    doi = "10.1088/1361-6382/ab5f2d",
    journal = "Class. Quant. Grav.",
    volume = "37",
    number = "4",
    pages = "045007",
    year = "2020"
}

@misc{GWTC-4zenodo,
author       = {{LIGO\textendash Virgo\textendash KAGRA Collaboration}},
  title        = {GWTC-4.0: Population Properties of Merging Compact Binaries},
  year         = {2025},
  publisher    = {Zenodo},
  doi          = {10.5281/zenodo.16911563},
  url          = {https://doi.org/10.5281/zenodo.16911563},
}

@article{Brandt1995,
    author = {Brandt, Niel and Podsiadlowski, Philipp},
    title = {The effects of high-velocity supernova kicks on the orbital properties and sky distributions of neutron-star binaries},
    journal = {Monthly Notices of the Royal Astronomical Society},
    volume = {274},
    number = {2},
    pages = {461-484},
    year = {1995},
    month = {05},
    issn = {0035-8711},
    doi = {10.1093/mnras/274.2.461},
    url = {https://doi.org/10.1093/mnras/274.2.461},
    eprint = {https://academic.oup.com/mnras/article-pdf/274/2/461/18161085/mnras274-0461.pdf},
}

@article{Kalogera1998,
doi = {10.1086/305086},
url = {https://doi.org/10.1086/305086},
year = {1998},
month = {jan},
publisher = {},
volume = {493},
number = {1},
pages = {368},
author = {Kalogera, Vassiliki},
title = {Formation of Low-Mass X-Ray Binaries. III. A New Formation Mechanism: Direct Supernova},
journal = {The Astrophysical Journal},
}

@inbook{Mapelli2021,
    author = "Mapelli, Michela",
    title = "{Formation Channels of Single and Binary Stellar-Mass Black Holes}",
    eprint = "2106.00699",
    archivePrefix = "arXiv",
    primaryClass = "astro-ph.HE",
    doi = "10.1007/978-981-15-4702-7_16-1",
    year = "2021"
}

@article{MandelBroekgaarden2021,
  author       = {Mandel, I. and Broekgaarden, F.~S.},
  title        = {Compact-object binary formation and evolution},
  journal      = {Living Reviews in Relativity},
  volume       = {25},
  number       = {1},
  pages        = {4},
  year         = {2022},
  doi          = {10.1007/s41114-021-00034-3},
  url          = {https://doi.org/10.1007/s41114-021-00034-3},
}

@ARTICLE{Renzo2019,
       author = {{Renzo}, M. and {Zapartas}, E. and {de Mink}, S.~E. and {G{\"o}tberg}, Y. and {Justham}, S. and {Farmer}, R.~J. and {Izzard}, R.~G. and {Toonen}, S. and {Sana}, H.},
       title = "{Massive runaway and walkaway stars. A study of the kinematical imprints of the physical processes governing the evolution and explosion of their binary progenitors}",
      journal = {\aap},
         year = 2019,
        month = apr,
       volume = {624},
          eid = {A66},
        pages = {A66},
          doi = {10.1051/0004-6361/201833297},
 primaryClass = {astro-ph.SR},
       adsurl = {https://ui.adsabs.harvard.edu/abs/2019A&A...624A..66R},
}

@article{giacobbo2018,
    author = {Giacobbo, Nicola and Mapelli, Michela},
    title = {The progenitors of compact-object binaries: impact of metallicity, common envelope and natal kicks},
    journal = {Monthly Notices of the Royal Astronomical Society},
    volume = {480},
    number = {2},
    pages = {2011-2030},
    year = {2018},
    month = {10},
    issn = {0035-8711},
    doi = {10.1093/mnras/sty1999},
    url = {https://doi.org/10.1093/mnras/sty1999},
    eprint = {https://academic.oup.com/mnras/article-pdf/480/2/2011/25440572/sty1999.pdf},
}

@article{yoon2010,
  title={Type Ib/c supernovae in binary systems. I. Evolution and properties of the progenitor stars},
  author={Yoon, S-C and Woosley, Stan E and Langer, Norbert},
  journal={The Astrophysical Journal},
  volume={725},
  number={1},
  pages={940--954},
  year={2010},
  publisher={The American Astronomical Society}
}

@ARTICLE{johnston1992,
       author = {{Johnston}, Simon and {Manchester}, R.~N. and {Lyne}, A.~G. and {Bailes}, M. and {Kaspi}, V.~M. and {Qiao}, Guojun and {D'Amico}, N.},
        title = "{PSR 1259-63: A Binary Radio Pulsar with a Be Star Companion}",
      journal = {\apjl},
         year = 1992,
        month = mar,
       volume = {387},
        pages = {L37},
          doi = {10.1086/186300},
       adsurl = {https://ui.adsabs.harvard.edu/abs/1992ApJ...387L..37J}
}

@ARTICLE{kaspi1994,
       author = {{Kaspi}, V.~M. and {Johnston}, S. and {Bell}, J.~F. and {Manchester}, R.~N. and {Bailes}, M. and {Bessell}, M. and {Lyne}, A.~G. and {D'Amico}, N.},
        title = "{A Massive Radio Pulsar Binary in the Small Magellanic Cloud}",
      journal = {\apjl},
         year = 1994,
        month = mar,
       volume = {423},
        pages = {L43},
          doi = {10.1086/187231},
       adsurl = {https://ui.adsabs.harvard.edu/abs/1994ApJ...423L..43K}
}

@ARTICLE{savonije1978,
       author = {{Savonije}, G.~J.},
        title = "{Roche-lobe overflow in X-ray binaries.}",
      journal = {\aap},
         year = 1978,
        month = jan,
       volume = {62},
       number = {3},
        pages = {317-338},
       adsurl = {https://ui.adsabs.harvard.edu/abs/1978A&A....62..317S}
}
\bibliographystyle{aasjournal}

\end{document}